\documentclass[]{article}

\renewcommand{\v}[1]{\ensuremath{\mathbf{#1}}}

\newcommand{\mc}{\mathcal}

\def\st{{\rm s.t.}}

\usepackage{algorithmic}
\usepackage{algorithm}
\usepackage{booktabs}
\usepackage[usenames,dvipsnames]{color}
\usepackage{fullpage,graphicx,amssymb,amsmath}
\usepackage{enumerate}
\usepackage{caption,subcaption}
\usepackage{empheq}

\usepackage[colorinlistoftodos,prependcaption,textsize=tiny]{todonotes}

\newcommand{\be}{\begin{enumerate}}
\newcommand{\ee}{\end{enumerate}}
\newtheorem{theorem}{Theorem}

\renewcommand{\Re}{\mathbb{R}} 
\newcommand{\Z}{\mathbb{Z}} 

\newcommand{\SNash}{\mc{S}^{\textrm{DC-Nash}}}
\newcommand{\SMCP}{\mc{S}^{\textrm{DC-MCP}}}

\newcommand{\fp}{f_p}

\newcommand{\xp}{\v{x}_p}
\newcommand{\xnotp}{\v{x}_{-p}}
\newcommand{\hatx}{\hat{\v{x}}}
\newcommand{\hatxp}{\hat{\v{x}}_p}
\newcommand{\hatxnotp}{\hat{\v{x}}_{-p}}
\newcommand{\xpr}{x_{pr}}
\newcommand{\Cp}{\mc{C}_p}
\newcommand{\Eqp}{\mc{E}_p}
\newcommand{\Ineqp}{\mc{I}_p}
\newcommand{\Intp}{\mc{D}_p}
\newcommand{\Xp}{\mc{X}_p}

\title{A Note on Solving Discretely-Constrained Nash-Cournot Games via Complementarity}
\author{Dimitri J. Papageorgiou$^*$, Francisco Trespalacios, Stuart Harwood \\
{\small Corporate Strategic Research}\\
{\small ExxonMobil Research and Engineering Company}\\
{\small 1545 Route 22 East, Annandale, NJ 08801 USA}\\
{\small \{dimitri.j.papageorgiou,francisco.trespalacios,stuart.harwood\}@exxonmobil.com} \\
{\small $^*$Corresponding author} \\
}

\begin{document}

\maketitle

\begin{abstract}
Discretely-constrained Nash-Cournot games have attracted attention as they arise in various competitive energy production settings in which players must make one or more discrete decisions.
Gabriel et al. \cite{Gabriel2013} claim that the set of equilibria to a discretely-constrained Nash-Cournot game coincides with the set of solutions to a corresponding discretely-constrained mixed complementarity problem. 
We show that this claim is false.  

\end{abstract}

\section{Introduction}

A Nash-Cournot game is a game-theoretical framework of imperfect competition in which multiple producers/players compete to optimize their individual objective functions, which also depend on other players' production decisions. 
Traditional (i.e., purely continuous) Nash-Cournot problems have been extensively studied and it is well known that they can be expressed either as nonlinear complementarity or variational inequality problems \cite{facchinei2007finite}. 
Discretely-constrained Nash-Cournot (DC-NC) games arise when a subset of a player's decisions are required to be discrete, for example, when a player must make a binary on/off decision.
Gabriel et al. \cite{Gabriel2013} approached discretely-constrained Nash-Cournot games by framing the problem as a discretely-constrained mixed complementarity problem (DC-MCP).  

We consider the same set up and, to the extent possible, the same notation as Gabriel et al. \cite{Gabriel2013}.
There are $N$ players indexed by $p \in \mc{P}=\{1,\dots,N\}$.  Player $p$ optimizes her cost function $f_p: \Re^{n} \mapsto \Re$ that depends on her decision vector $\v{x}_p \in \Re^{n_p}$ and the vector $\xnotp = (\v{x}_1, \dots, \v{x}_{p-1}, \v{x}_{p+1}, \dots, \v{x}_N)$ denoting the decisions of all other players besides player $p$. Here, $n = \sum_{p \in \mc{P}} n_p$. Specifically, we assume that player $p$ solves the following discretely-constrained optimization problem parameterized by $\xnotp$:
\begin{subequations} \label{model:gabriel_nc_player}
\begin{alignat}{4}
\fp^*(\xnotp) =
\min_{\xp}~~& \fp(\xp,\xnotp) & & \qquad {\color{red} [\textrm{dual vars}]} \\ 
\st~~& g_{pj}(\xp) \leq 0 & & \qquad {\color{red} [\lambda_{pj} \geq 0]} \qquad \forall j \in \Ineqp \label{eq:dcnc_inequalities} \\
	 & h_{pk}(\xp) = 0 & & \qquad {\color{red} [\gamma_{pk} \in \Re]} \qquad \forall k \in \Eqp \label{eq:dcnc_equalities}  \\
	 & \xp \geq \v{0} & & \label{eq:dcnc_nonnegativity}  \\
	 & \xpr \in \Z_+ & & \qquad \qquad \qquad \qquad \forall r \in \Intp~, \label{eq:dcnc_integers}  
\end{alignat}
\end{subequations}
where $\Ineqp$, $\Eqp$, and $\Intp$ denote the set of inequalities, equalities, and integer variables for player $p \in \mc{P}$.
Let $\Xp = \{ \xp \in \Re^{n_p} : \eqref{eq:dcnc_inequalities}, \eqref{eq:dcnc_equalities}, \eqref{eq:dcnc_nonnegativity}, \eqref{eq:dcnc_integers} \}$ denote the discretely-constrained feasible region for player $p \in \mc{P}$ and let $\Cp = \{ \xp \in \Re^{n_p} : \eqref{eq:dcnc_inequalities}, \eqref{eq:dcnc_equalities}, \eqref{eq:dcnc_nonnegativity} \}$ denote the continuous relaxation of $\Xp$.
A vector $\hatx$ is called a Nash equilibrium of this DC-NC game if $\hatxp \in \Xp$ for all $p \in \mc{P}$ and 
\begin{equation} \label{game:DCNC}
\fp(\hatxp,\hatxnotp) \leq \fp(\xp,\hatxnotp), \qquad \forall p \in \mc{P}, \xp \in \Xp~.
\end{equation}

Gabriel et al. \cite{Gabriel2013} approach convex DC-NC games, i.e., games in which the continuous relaxation of each player's optimization problem is a convex optimization problem, by applying the following four-step procedure: 1) relax the integrality constraints for each player; 2) write the KKT conditions for each player; 3) re-impose the integrality constraints; 4) solve the resulting DC-MCP.
More concretely, since KKT conditions are neither necessary nor sufficient for a discrete optimization problem, 
Gabriel et al. \cite{Gabriel2013} attempt to find the set of Nash equilibria to \eqref{game:DCNC} by appealing to the continuous relaxation of each player's parametric optimization problem: 
\begin{equation} \label{model:Nash_player_continuous_relaxation}
\min \Big\{ \fp(\xp,\xnotp) : \xp \in \Cp \Big\}~.
\end{equation}
Assume that the functions $\fp(\cdot,\xnotp)$ are convex and a constraint qualification for the continuous relaxation $\Cp$ holds. Then, the KKT conditions for player $p$'s relaxed problem \eqref{model:Nash_player_continuous_relaxation} are to find $\xp \in \Re^{n_p}$, $\lambda_p \in \Re^{|\Ineqp|}$, $\gamma_p \in \Re^{|\Eqp|}$ such that
\begin{subequations} \label{model:gabriel_kkt}
\begin{alignat}{4}
& & \v{0} \leq \nabla_{\xp} \fp(\xp,\xnotp) 
+ \sum_{j \in \Ineqp} \lambda_{pj} \nabla g_{pj}(\xp) 
+ \sum_{k \in \Eqp} \gamma_{pk} \nabla h_{pk}(\xp)  
\perp \xp \geq \v{0} & \\ 
& & 0 \leq -g_{pj}(\xp,\xnotp) \perp \lambda_{pj} \geq 0 \qquad \forall j \in \Ineqp & \\
& 	 & h_{pk}(\xp,\xnotp) = 0~,~\gamma_{pk} \in \Re \qquad \forall k \in \Eqp  & 
\end{alignat}
\end{subequations}
Gabriel et al. \cite{Gabriel2013} (p.313) then write:
\begin{quote}
``An interesting question is whether the set of $\xp$ that solves \eqref{model:gabriel_kkt}, but with the discrete restrictions for $\xpr \in \Z_+$ for $r \in \Intp$, corresponds to the solution set of the original problem \eqref{game:DCNC}.  The next result shows that this correspondence is correct."
\end{quote}
\begin{theorem}[Theorem 3 in \cite{Gabriel2013}]
Let $\SNash$ be the set of solutions to the discretely-constrained Nash-Cournot game \eqref{game:DCNC} and $\SMCP$ be the set of solutions to \eqref{model:gabriel_kkt} for which $\xpr \in \Z_+$ for $r \in \Intp$. Then, $\SNash = \SMCP$. 
\end{theorem}

\section{Counterexamples}

We now provide two simple discretely-constrained Nash-Cournot duopoly games (i.e., $\mc{P} = \{1,2\}$) for which one or more equilibria exist to \eqref{game:DCNC}, but the complementarity conditions coupled with integrality restrictions are either 1) empty, or 2) non-empty, but a strict subset of the true set of equilibria.  
In both examples, because each player controls a single decision variable, we index player $p$'s decision variable as $x_p$ rather than $x_{p1}$. 

\subsection{``Linear" players with weak continuous relaxations}

Consider the simple Nash-Cournot duopoly game with the following symmetric payoff matrix: 
\begin{center}
\begin{tabular}{c|cc}
		  & $x_2 = 0$ & $x_2 = 1$ \\
\hline		  
$x_1 = 0$ &  0 & -1 \\
$x_1 = 1$ & -1 & -2 \\
\end{tabular}
\end{center}
Here each player can take a discrete (binary) action with the unique equilibrium being $x_1=x_2=1$, i.e., each player chooses action 1 for a (minimum) payoff of -2, which is obviously a dominant strategy for each player.
We now translate this DC-NC game into an optimization framework. Suppose player $p \in \{1,2\}$ solves the following problem:
\begin{equation}
\fp^*(x_{-p}) = \min \Big\{ -x_p - x_{-p} : x_p \in [0,1+\epsilon] \cap \Z \Big\}~,
\end{equation}
where $\epsilon > 0$ and $\Z$ is the set of integers.
The corresponding KKT optimality conditions are 
\begin{subequations} \label{counterexample:kkt}
\begin{alignat}{4}
& 0 \leq \lambda_p - 1 & & \perp x_p \geq 0 \qquad \forall p \label{counterexample:kkt_obj} \\ 
& 0 \leq 1+\epsilon - x_p & & \perp \lambda_p \geq 0 \qquad \forall p \label{counterexample:kkt_var_upper_bound}
\end{alignat}
\end{subequations}
We now plug in the unique equilibrium solution $x_1=x_2=1$. Complementarity conditions \eqref{counterexample:kkt_obj} imply that $\lambda_p = 1$, while conditions \eqref{counterexample:kkt_var_upper_bound} imply that $\lambda_p = 0$.  This contradiction reveals that the unique equilibrium solution $x_1=x_2=1$ is not in $\SMCP$, i.e. $\emptyset = \SMCP \subset \SNash \neq \emptyset$.

\subsection{``Quadratic" players with tight continuous relaxations}

In this example, the continuous relaxation for each player is tight.
Consider the payoff matrix
\begin{center}
\begin{tabular}{c|cc}
		  & $x_2 = 0$ & $x_2 = 1$ \\
\hline		  
$x_1 = 0$ & $(0,0)$ & $(9,9)$ \\
$x_1 = 1$ & $(4,4)$ & $(1,1-\delta)$ \\
\end{tabular}
\end{center}
For $\delta > -3$, there are two equilibria in pure strategies: $(x_1,x_2) = (0,0)$ and $(x_1,x_2) = (1,1)$.

This corresponds to player 1 solving the following convex quadratic problem (as a function of $x_2$): 
\begin{subequations} \label{model:counterexample_djp1_p1}
\begin{alignat}{4}
f_1^*(x_2) = \min_{x_1}~~& (2x_1 - 3x_2)^2 & & \\ 
\st~~& x_1 - 1 \leq 0 & & \\
	 & -x_1 \leq 0  & & 
\end{alignat}
\end{subequations}
Meanwhile, player 2 solves a similar convex quadratic problem (as a function of $x_1$): 
\begin{subequations} \label{model:counterexample_djp1_p2}
\begin{alignat}{4}
f_2^*(x_1) = \min_{x_2}~~& (2x_1 - 3x_2)^2 - \delta x_1 x_2 & & \\ 
\st~~& x_2 - 1 \leq 0 & & \\
	 & -x_2 \leq 0  & & 
\end{alignat}
\end{subequations}
Note that $\fp(\cdot,\xnotp)$ are convex functions and a constraint qualification holds.

The KKT conditions \eqref{model:gabriel_kkt} become
\begin{subequations} \label{counterexample_djp1:kkt}
\begin{alignat}{4}
& 0 \leq 4(2x_1 - 3x_2) + \lambda_1  & & \perp x_1 \geq 0  \label{counterexample_djp1:kkt_obj_p1} \\ 
& 0 \leq -6(2x_1 - 3x_2) -\delta x_1 + \lambda_2  & & \perp x_2 \geq 0 \label{counterexample_djp1:kkt_obj_p2} \\ 
& 0 \leq 1 - x_p & & \perp \lambda_p \geq 0 \qquad \forall p \label{counterexample_djp1:kkt_var_upper_bound}
\end{alignat}
\end{subequations}

Assume $\delta > -3$. It is straightforward to verify that $x_p = \lambda_p = 0$ for all $p$ satisfy the complementarity conditions~\eqref{counterexample_djp1:kkt}. The situation is different for $(x_1,x_2) = (1,1)$. Condition~\eqref{counterexample_djp1:kkt_obj_p1} implies that $\lambda_1 = 4$, while condition~\eqref{counterexample_djp1:kkt_var_upper_bound} implies that $\lambda_p \geq 0$ for all $p$. However, condition~\eqref{counterexample_djp1:kkt_obj_p2} implies that $\lambda_2 = -6+\delta$.  Thus, for $\delta \in (-3,6)$, the complementarity approach fails to recognize $(x_1,x_2) = (1,1)$ as an equilibrium.  It it tempting to argue that when $\delta \in (-3,1]$, this omission is not a concern because $(x_1,x_2) = (0,0)$ is the preferred equilibrium (i.e., the global minimizer for both players). However, for $\delta > 1$, player 2's global minimizer is $(x_1,x_2) = (1,1)$ with a payoff of $1-\delta$ and, for $\delta \in (1,6)$, the complementarity approach does not ``see" this solution as an equilibrium. In short, this example shows that, not only can the complementarity approach fail to find all equilibria to a DC-NC game, it is not guaranteed to find global optima for each player when it does return an equilibrium.

Note that one can obtain a similar result (counterexample) by replacing the $L2$ term $(2x_1-3x_2)^2$ with the $L1$ term $|2x_1-3x_2|$ so that each player solves a linear optimization problem instead of a convex quadratic one.   

%
%

%

\section{Resolution}

For completeness, the correct version of Theorem 3 in Gabriel et al. is
\begin{theorem}
Let $\SNash$ be the set of solutions to the discretely-constrained Nash-Cournot game \eqref{game:DCNC} and $\SMCP$ be the set of solutions to \eqref{model:gabriel_kkt} for which $\xpr \in \Z_+$ for $r \in \Intp$. Then, $\SMCP \subseteq \SNash$ and there exist cases when $\SMCP \subsetneq \SNash$. 
\end{theorem}
Finally, note that the heuristic proposed by Gabriel et al. to solve the DC-NC game \eqref{game:DCNC} is still valid. 

\small
\bibliographystyle{abbrv}
\bibliography{gma_refs}

\end{document}